\documentclass[aps,prb,twocolumn,superscriptaddress,floatfix,nofootinbib]{revtex4-2}

\makeatletter
\AtBeginDocument{\if@filesw\immediate\write\@auxout{\string\bibstyle{apsrev4-2}}\fi}
\makeatother

\usepackage{amsmath}
\usepackage{amssymb}
\usepackage{bm}
\usepackage{graphicx}
\usepackage{color}
\usepackage{xcolor}
\usepackage{siunitx}
\usepackage{booktabs}
\usepackage[colorlinks=true,allcolors=blue]{hyperref}
\usepackage{soul}

\definecolor{lightpink}{RGB}{255, 182, 193}
\sethlcolor{lightpink}
\definecolor{pink}{RGB}{255, 105, 180}

\newcommand{\NO}{\mathrm{NO}^{\bullet}}
\newcommand{\Osup}{\mathrm{O}_2^{\bullet-}}
\newcommand{\Tone}{T_1}

\newcommand{\kB}{k_{\mathrm{B}}}
\newcommand{\uM}{\,\mu\mathrm{M}}

\begin{document}

\title{Phonon-limited detection thresholds for genetically encoded
fluorescent-protein spin-qubit relaxometry of neural radicals}

\author{Parul Raghuvanshi}
\affiliation{
Indian Institute of Technology Roorkee, Roorkee 247667, India.}\author{Sagnik Ganguly}
\affiliation{
Indian Institute of Engineering Science and Technology, Shibpur, Howrah 711103, India.}
\author{Sharika E}
\affiliation{
Indian Institute of Technology Roorkee, Roorkee 247667, India.}
\author{Mohana Priya T.}
\affiliation{
Indian Institute of Technology (ISM) Dhanbad, Dhanbad 826004, India.}
\author{Vishvendra S. Poonia}
\email{vishvendra@ece.iitr.ac.in}
\affiliation{
Indian Institute of Technology Roorkee, Roorkee 247667, India.}


\date{\today}

\begin{abstract}
The demonstration that enhanced yellow fluorescent protein hosts an
optically addressable spin-1 qubit in its metastable triplet state raises
the prospect of genetically encoded quantum sensing at molecular length
scales. We develop a detection-limit theory for using this
fluorescent-protein spin qubit (FPSQ) to sense paramagnetic neural
signaling radicals by spin relaxometry. We derive the transition-resolved
Redfield relaxation matrix of the zero-field-split triplet coupled to a
diffusing radical bath, establish the regime in which it collapses to a
single exponential, and validate it against Lindblad simulations and
nitrogen-vacancy benchmarks. Propagating the effects of photon shot noise, photobleaching-grounded photon budget, and finite measurement bandwidth,
we find that the native room-temperature sensor falls short of
physiological sensitivity by six to eight orders of magnitude with the bottleneck being the phonon-limited intrinsic $\Tone$. Analyzing the
underlying direct and two-phonon Raman processes, we show that
room-temperature relaxation is Raman-dominated by $\sim\!720\!:\!1$ and
that, because the Raman coefficient scales as $v^{-10}$ with sound
velocity, a $\sim\!2\times$ stiffening of the chromophore environment
recovers $\Tone\sim\SI{100}{\micro\second}$, sufficient for micromolar
sensing. Nanomolar sensing is obstructed by a direct-process ceiling of
\SI{79}{\micro\second} that vibronic decoupling alone cannot breach. We
obtain quantitative design rules, identify photon yield as a co-equal
bottleneck, and propose a frequency-resolved protocol for chemical
specificity.
\end{abstract}

\maketitle

\section{Introduction}\label{sec:intro}

Nitric oxide and reactive oxygen species (ROS) are among the most
consequential small-molecule signals in the nervous system. Nitric oxide,
produced by neuronal nitric oxide synthase (nNOS) in response to
NMDA-receptor activation, is a freely diffusing retrograde messenger that
regulates synaptic plasticity, cerebral blood flow, and neuroendocrine
function~\cite{garthwaite2008,picon2019,Toda2009}. Superoxide ($\Osup$) and its downstream
partners act both as damaging byproducts of oxidative metabolism and as
bona fide signaling species with defined roles in neuronal excitability and
long-term potentiation~\cite{eisenstein2020,huddleston2008,klann1998}. Because these radicals are
generated at specific subcellular sites, such as synaptic densities, mitochondria,
membrane microdomains and consumed on submicron length scales, a sensor
reporting their local concentration with molecular spatial resolution and
genetically defined placement would address a long-standing measurement
gap~\cite{wardman2007}. Electrochemical microelectrodes and small-molecule fluorescent
indicators either lack nanoscale placement or perturb the redox environment
they measure, and none reports the paramagnetism of the radical directly~\cite{wardman2007,privett2010,trouillon2013,wang2020,wang2006}.

Spin relaxometry offers such a direct readout. A paramagnetic radical
carries an electron spin whose fluctuating dipolar field shortens the
longitudinal relaxation time $\Tone$ of a nearby sensor spin.Therefore measuring
$\Tone$ provides a measure of the local density of unpaired
electrons ~\cite{colehollenberg2009,steinert2013}. This principle underlies
a mature body of work with nitrogen-vacancy (NV) centers in nanodiamonds,
used to detect gadolinium spin labels, ferritin, and free radicals
including $\NO$ and superoxide in and around living
cells~\cite{tetienne2013,steinert2013,rendler2017,sigaeva2022,peronamartinez2020}.
Nanodiamond relaxometry is nonetheless constrained by delivery , as the diamond
is an exogenous particle of finite size, cannot be targeted to a chosen
protein with nanometer precision, and typically sits \SIrange{10}{20}{nm}~\cite{sigaeva2022}
from the radical source across its own crystal and surface-passivation
layers.

\begin{figure*}[t]
\centering
\includegraphics[width=\textwidth]{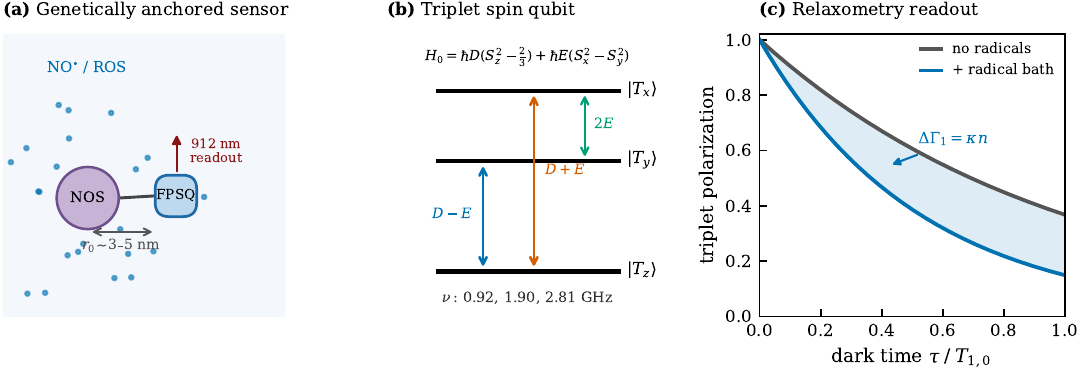}
\caption{\textbf{Proposed FPSQ relaxometry concept.}
(a) A genetically anchored fluorescent protein spin qubit (FPSQ) is placed close to a radical-producing enzyme and read out optically at 912 nm. 
(b) The EYFP triplet acts as a spin-1 qubit with three zero-field ODMR transitions at $2E$, $D-E$, and $D+E$. 
(c) A fluctuating radical bath adds a longitudinal relaxation channel, $\Delta\Gamma_1 = \kappa n$, which shortens the triplet polarization decay measured after a dark time $\tau$. This change is used to estimate the local radical density.}
\label{fig:concept}
\end{figure*}

The demonstration by Feder \emph{et al.}that enhanced yellow fluorescent
protein (EYFP) behaves as an optically addressable spin-1
qubit~\cite{feder2025fluorescent} changes the delivery problem qualitatively. The
qubit lives in the metastable triplet state ($T_1$) of the chromophore; it
is initialized by \SI{488}{nm} excitation through spin-selective
intersystem crossing, coherently manipulated by microwaves, and read out
optically through an optically activated delayed-fluorescence (OADF) scheme
in which a \SI{912}{nm} pulse drives $T_1\!\to\!T_2\!\to\!S_1$ and the
resulting delayed fluorescence, temporally separated from prompt
autofluorescence, encodes the spin state. The triplet Hamiltonian is that
of a spin-1 with zero-field splitting,
\begin{equation}
H_0 = \hbar D\!\left(S_z^2 - \tfrac{2}{3}\right)
      + \hbar E\!\left(S_x^2 - S_y^2\right)
      - \hbar\gamma_e\,\mathbf{S}\!\cdot\!\mathbf{B},
\label{eq:H0}
\end{equation}
with measured $D = 2\pi\times\SI{2.356}{GHz}$ and
$E = 2\pi\times\SI{0.458}{GHz}$, giving three zero-field optically detected
magnetic resonance (ODMR) transitions at $D\!-\!E\approx\SI{1.90}{GHz}$,
$D\!+\!E\approx\SI{2.81}{GHz}$, and $2E\approx\SI{0.92}{GHz}$. Because the
qubit is a genetically encoded protein it can be fused to a target of
choice , placing the sensor spin within the $\sim\!3$ nm diameter of the
$\beta$-barrel of a radical-producing enzyme\cite{feder2025fluorescent}, and its sensing properties can be further optimized by directed evolution~\cite{feder2025fluorescent,mann2026optically} [Fig.~\ref{fig:concept}].

This paper asks a single quantitative question: \emph{can a genetically
anchored FPSQ detect physiological neural radicals by relaxometry, and if
not, what specific material property must change?} We answer with a
detection-limit theory whose central conclusion is a statement about
spin-phonon physics rather than about biology. Section~\ref{sec:theory}
derives the transition-resolved spin-1 Redfield relaxation matrix, shows
when it reduces to a single exponential, and validates it numerically.
Section~\ref{sec:phonon} analyzes the intrinsic $\Tone$  by decomposing the
measured $AT+BT^7$ law into its direct and two-phonon Raman contributions,
shows that room-temperature relaxation is overwhelmingly Raman-limited, and
uses Debye-model scaling to convert the required $\Tone$ improvement into a
concrete target for the stiffness and vibronic coupling of the chromophore
pocket. Section~\ref{sec:framework} propagates shot noise, a photobleaching-grounded photon budget, and finite bandwidth to a minimum
detectable concentration. Section~\ref{sec:results} presents the
detection-limit map, the sensitivity-bandwidth tradeoff, the robustness of
the thresholds, the genetic-anchoring advantage, and a frequency-resolved
species-discrimination protocol. Section~\ref{sec:discussion} states the
verdict, the design rules, testable predictions, and limitations.

Three results emerge that we believe are of interest beyond the specific
platform. 
First, the room-temperature relaxation of this molecular qubit is Raman-dominated by nearly three orders of magnitude, so the seemingly hopeless requirement of a $10^3$--$10^4$ increase in $\Tone$ maps onto a $\sim\!2\times$ change in effective stiffness through the steep $v^{-10}$ scaling of the Raman coefficient. Second, suppressing the vibronic coupling
alone cannot succeed because a direct-process ceiling of $\sim\!\SI{79}{\micro
\second}$ at \SI{295}{K} caps the achievable $\Tone$ unless the phonon
spectrum itself is stiffened. Third, the photon yield per molecule is a
co-equal bottleneck to $\Tone$, and single-cell operation at physiological
concentrations requires both to be solved together.

\section{Spin-1 relaxometry theory}\label{sec:theory}

{\subsection{Dipolar coupling and the spin-1 relaxation matrix}
\label{sec:dipolar}

A radical of electron spin $S_R$ at position $\mathbf{r}$ relative to the
qubit produces a magnetic dipole field. The qubit-radical dipolar
interaction has the characteristic coupling frequency
\begin{equation}
b(r) = \frac{\mu_0}{4\pi}\,\frac{\gamma_e^2 \hbar}{r^3}
     = \frac{\mu_0}{4\pi}\,\frac{g^2\mu_B^2}{\hbar\, r^3},
\label{eq:b}
\end{equation}
where $\gamma_e = g\mu_B/\hbar$. For two electron spins this evaluates to
$b/2\pi = \SI{52.0}{MHz}$ at $r=\SI{1}{nm}$, falling as $r^{-3}$ to
\SI{6.5}{MHz} at \SI{2}{nm} and \SI{1.9}{MHz} at \SI{3}{nm}. When the
radical reorients or diffuses, the transverse components of this field
fluctuate, treating them as a stationary Gaussian process with correlation
time $\tau_c$ gives the Lorentzian noise spectrum
\begin{equation}
J(\omega) = \frac{\tau_c}{1+(\omega\tau_c)^2}.
\label{eq:J}
\end{equation}

The sensor is a spin-1, not a two-level system, and this deserves care. The
zero-field-split triplet has three nondegenerate eigenstates
$\{|T_x\rangle,|T_y\rangle,|T_z\rangle\}$, the eigenstates of
Eq.~\eqref{eq:H0} at $B=0$, separated by the three unequal frequencies
$\omega_{xz}=D\!+\!E$, $\omega_{yz}=D\!-\!E$, and $\omega_{xy}=2E$.
Second-order (Redfield/Bloch--Wangsness) perturbation
theory~\cite{bpp1948,redfield1957,abragam1961,slichter1990} in the
fluctuating field $\mathbf{B}_{\rm loc}(t)$ generated by the radical gives a
rate matrix rather than a single rate. The population-transfer rate
between eigenstates $|i\rangle$ and $|j\rangle$ is  given by
\begin{equation}
W_{ij} = \gamma_e^2 \sum_{\mu\in\{x,y,z\}}
         \big|\langle i|S_\mu|j\rangle\big|^2\,
         S_{\mu\mu}(\omega_{ij}),
\label{eq:Wij}
\end{equation}
where $S_{\mu\mu}(\omega) = \langle B_\mu^2\rangle J(\omega)$ is the
one-sided noise power of the $\mu$ component of the local field at the
transition frequency $\omega_{ij}=|E_i-E_j|/\hbar$. For the ZFS eigenbasis
one finds the selection structure that each pair $(i,j)$ is connected by
exactly one cartesian spin operator with unit matrix element,
$\sum_\mu|\langle i|S_\mu|j\rangle|^2 = 1$, so that each transition is
driven by an orthogonal component of the field noise sampled at its own
frequency i.e., $W_{xz}\propto J(\omega_{xz})$, $W_{yz}\propto J(\omega_{yz})$,
and $W_{xy}\propto J(\omega_{xy})$.

The populations obey $\dot{P}_i = \sum_{j\neq i}(W_{ij}P_j - W_{ji}P_i)$. In
the high-temperature limit relevant here ($\hbar\omega_{ij}\ll \kB T$, i.e.\
$\SI{2.8}{GHz}\ll\SI{6}{THz}$ at \SI{295}{K}), the up and down rates are
equal and the stationary state is fully mixed, with $P_i=1/3$. Initializing the
polarized state $|T_z\rangle$ and following the measured population, the
initial decay obeys
\begin{equation}
\Gamma_1^{\rm eff} \equiv
-\frac{1}{P_{T_z}-\tfrac13}\frac{dP_{T_z}}{dt}\bigg|_{t=0}
= \tfrac{3}{2}\left(W_{xz}+W_{yz}\right),
\label{eq:Gammaeff}
\end{equation}
i.e.\ the $|T_z\rangle$ polarization relaxes through the two channels that
leave it, weighted by the $2/3$ initial polarization excess. The
$|T_x\rangle\!\leftrightarrow\!|T_y\rangle$ channel does not enter the
initial decay of $|T_z\rangle$ but controls the approach to equilibrium at
longer times.}

\subsection{When the single-exponential description is exact}
\label{sec:singleexp}

\begin{figure*}[t]
\centering
\includegraphics[width=\textwidth]{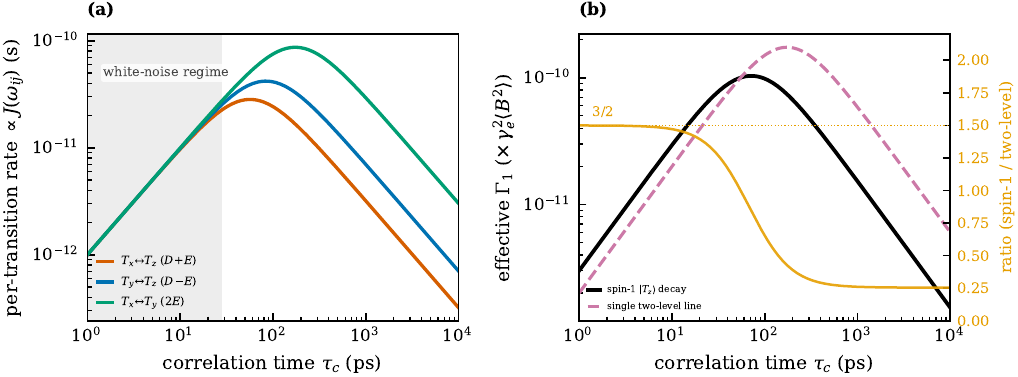}
\caption{\textbf{Transition-resolved spin-1 relaxation.} 
(a) Population transfer rates for the three zero-field transitions, each driven by an orthogonal component of the dipolar field noise sampled at its own frequency $\omega_{ij}$. In the white-noise regime (shaded, $\tau_c\!\lesssim\!\SI{57}{ps}$), the three rates are degenerate. Beyond this regime, they split, making the decay multi-exponential and the readout transition a design variable. 
(b) Effective $|T_z\rangle$ polarization decay rate 
versus correlation time $\tau_c$
compared with a single two-level approximation. In the white-noise regime, the spin-1 relaxation rate is $3/2$ times the two-level estimate, relevant to NO• and superoxide, confirming the use of  single-exponential detection-limit analysis for the target analytes.
}
\label{fig:spin1}
\end{figure*}

Equation~\eqref{eq:Gammaeff} shows that a single relaxation rate is not
generally sufficient because the three $J(\omega_{ij})$ differ, the decay is in
principle multi-exponential, and the rate depends on which sublevel is
initialized and read out. Two limits clarify matters [Fig.~\ref{fig:spin1}].

In the white-noise limit $\omega_{ij}\tau_c\ll1$ for all three
transitions i.e.\ $\tau_c \ll 1/\omega_{xz} \approx \SI{57}{ps}$ , the
spectral density is flat, $J(\omega_{ij})\to\tau_c$, all three transition
rates become degenerate, and the population dynamics collapse rigorously to
a single exponential with $\Gamma_1^{\rm eff}=3W$,
$W=\gamma_e^2\langle B_\mu^2\rangle\tau_c$. This is precisely the regime of
the fast reactive radicals of interest such as $\NO$ has
$\tau_c\sim\SIrange{1}{10}{ps}$ and superoxide
$\tau_c\sim\SIrange{10}{100}{ps}$ (Sec.~\ref{sec:tauc}). For the
target analytes, therefore, the single-exponential model is not an
approximation but an exact consequence of the noise being white across the
ZFS manifold. This resolves what would otherwise be a significant
ambiguity in interpreting relaxometry data on a spin-1 sensor.

Outside that limit, for slowly relaxing radicals such as nitroxide spin
labels with $\tau_c\gtrsim\SI{1}{ns}$, the three rates split by up to an
order of magnitude, the decay becomes visibly multi-exponential, and the
readout transition becomes a design variable [Fig.~\ref{fig:spin1}(a)].
This splitting is not a nuisance. Instead, it is the physical origin of the
frequency-resolved chemical specificity developed in Sec.~\ref{sec:discrim}.
Figure~\ref{fig:spin1}(b) quantifies the departure from a naive two-level
estimate, showing the $\tfrac32$ enhancement in the white-noise limit and
the crossover beyond it.

Because the target analytes sit in the white-noise regime, we use the
compact angle-averaged single-radical rate given as
\begin{equation}
\Gamma_1^{\mathrm{single}}(r) =
\tfrac{8}{3}\,S_R(S_R+1)\,b(r)^2\, J(\omega_0),
\label{eq:gamma1single}
\end{equation}
which for $S_R=\tfrac12$ has prefactor $2$, with the spin-1 structure
verified to contribute only the $\mathcal{O}(1)$ factor derived above. All
thresholds below are order-of-magnitude statements insensitive to this
factor. The weak-coupling (motional-narrowing) condition $b(r)\,\tau_c \ll
1$ is very well satisfied: at $r=\SI{2}{nm}$ with $\tau_c=\SI{5}{ps}$,
$b\tau_c \sim 2\times10^{-4}$.

\subsection{Bath integral and relaxivity}\label{sec:bath}

For a uniform bath of radicals at number density $n$ with closest approach
$a$, integrating Eq.~\eqref{eq:gamma1single} over the bath with a
distance-independent correlation time gives
\begin{align}
\Gamma_1^{\mathrm{bath}}
&= n\!\int_a^{\infty}\!\! \tfrac{8}{3}S_R(S_R+1)\,b(r)^2 J(\omega_0)\,
   4\pi r^2\, dr \nonumber\\
&= \underbrace{\tfrac{32\pi}{9}\,S_R(S_R+1)\,a^3\,b(a)^2\,J(\omega_0)}_{\displaystyle
   \kappa}\; n,
\label{eq:kappa}
\end{align}
defining the relaxivity $\kappa \equiv d\Gamma_1/dn$ (units
$\si{s^{-1}.m^3}$). The $r^{-6}$ weighting makes the integral converge
sharply at the inner cutoff, so $\kappa$ is dominated by the closest
radicals and scales as $a^{-3}$. For a spin-$\tfrac12$ radical at
$a=\SI{2}{nm}$ with $\tau_c=\SI{10}{ps}$ on the lowest ODMR transition,
$\kappa$ corresponds to a molar relaxivity $r_1 \approx \SI{0.67}{s^{-1}}$
per micromolar ($\kappa\approx\SI{1.1e-18}{s^{-1}.m^3}$). In the
white-noise regime $\kappa$ is linear in $\tau_c$.

The $a^{-3}$ scaling makes the inner cutoff physically load-bearing rather
than a formal device. It represents the distance of closest approach
between the radical and the chromophore's unpaired spin density, set by the
$\beta$-barrel wall and the excluded volume of the two molecules, so it cannot
be taken below $\sim\!1$ nm without abandoning both the point-dipole
description of the delocalized chromophore spin density and the
weak-coupling condition. We treat $a$ as a bounded design parameter and
report the sensitivity of all conclusions to it (Sec.~\ref{sec:robust}).

\subsection{Correlation time and the near-field regime}\label{sec:tauc}

\begin{figure*}[t]
\centering
\includegraphics[width=\textwidth]{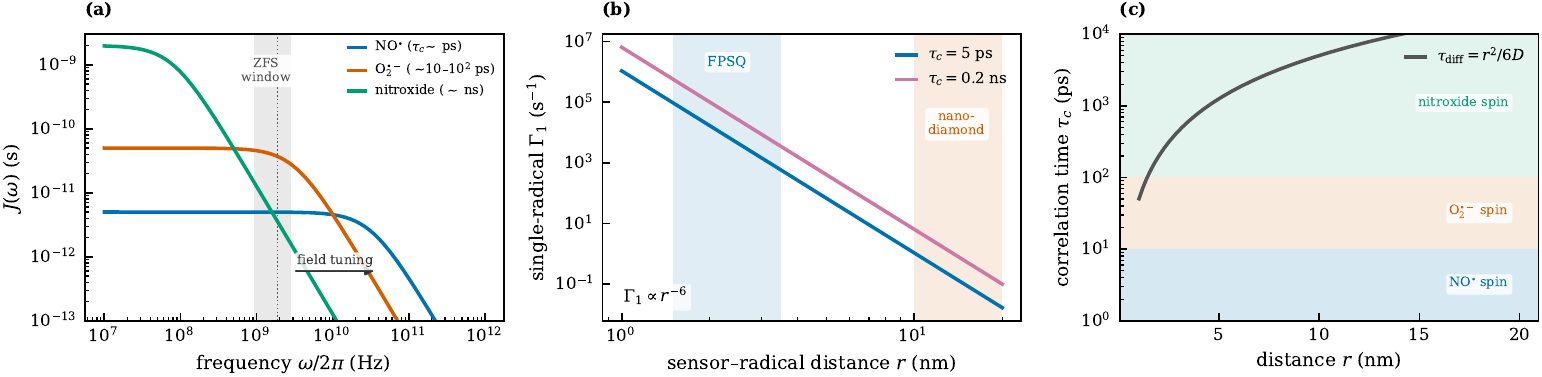}
\caption{\textbf{Relaxometry physics.} (a) Lorentzian noise spectral
density $J(\omega)$ for different correlation times, representative of $\NO$ (\si{ps}),superoxide (tens of \si{ps}), and nitroxide labels (\si{ns}). The zero-field ODMR window and the range accessible by field tuning are
indicated. (b) Single-radical relaxation rate as a function of sensor-radical distance, scaling as $r^{-6}$.The shaded regions indicate the distance ranges for the anchored FPSQ and nanodiamond. (c) Correlation time as a function of distance. The diffusion time $\tau_{\mathrm{diff}}=r^2/6D$ increases with distance and competes with
the intrinsic spin-relaxation time of each radical, shown as shaded
bands, with the shorter timescale determining the effective correlation
time.}
\label{fig:physics}
\end{figure*}

The correlation time entering $J(\omega_0)$ is set by whichever process
decorrelates the dipolar field fastest either through translational diffusion of the
radical past the sensor, or through the radical's own electron-spin relaxation:
\begin{equation}
\tau_c^{-1} = \tau_{\mathrm{diff}}^{-1} + \tau_{\mathrm{spin}}^{-1},
\qquad
\tau_{\mathrm{diff}} = \frac{r^2}{6D}.
\label{eq:tauc}
\end{equation}
For nitric oxide in aqueous medium at \SI{37}{\celsius},
$D\approx\SI{3300}{\micro\meter\squared\per\second}$~\cite{malinski1993,zachariadeen2005},
so $\tau_{\mathrm{diff}}\approx\SI{0.2}{ns}$ at $r=\SI{2}{nm}$. The
electron-spin relaxation of $\NO$ in solution is by contrast extremely
fast, with dissolved $\NO$ is essentially unobservable by EPR at room
temperature because its $^2\Pi_{1/2}$/$^2\Pi_{3/2}$ near-degeneracy provides
an efficient spin--rotation relaxation channel, giving $\tau_{\rm spin}$ of
order picoseconds. For $\NO$ the effective $\tau_c$ is therefore
spin-relaxation-dominated and short, ranging from \SIrange{1}{10}{ps}. Superoxide relaxes
more slowly through $g$-anisotropy modulation,
$\tau_c\sim\SIrange{10}{100}{ps}$, and tumbling nitroxide labels extend to
\SIrange{0.1}{10}{ns}. This spread is genuinely uncertain for the reactive
species and we treat it as a swept parameter throughout
[Fig.~\ref{fig:physics}(a)]. It also serves as the physical handle on chemical specificity (Sec.~\ref{sec:discrim}).

Two consequences deserve emphasis. First, because $\tau_{\rm diff}$ grows as
$r^2$, a sensor a few nanometers from the source samples the noise at a
correlation time far shorter than a nanodiamond \SIrange{10}{20}{nm} away.
Combined with the $r^{-6}$ coupling, this is a near-field advantage intrinsic
to genetic anchoring [Fig.~\ref{fig:physics}(b,c)]. Second, when $\tau_c$ is
set by fast spin relaxation, the noise is white across the entire ODMR band,
so $J(\omega_0)\approx\tau_c$, independent of which transition is probed;
frequency resolution then requires tuning $\omega_0$ well above the
zero-field splitting with an applied field (Sec.~\ref{sec:discrim}).

\subsection{Numerical validation}\label{sec:validation}

We validated Eqs.~\eqref{eq:Wij}--\eqref{eq:kappa} against independent
numerics in three ways with details provided in \textcolor{blue}{\href{run:Supplementary_Material.pdf}{Supplementary Material}}. First, we simulated the full spin-1 qubit of
Eq.~\eqref{eq:H0} dipolar-coupled to an explicit spin-$\tfrac12$ radical
whose electron spin undergoes stochastic reorientation, propagating the
joint density matrix with a Lindblad master
equation~\cite{qutip2012,qutip2013} and extracting the qubit longitudinal
relaxation rate. The simulated rate reproduces the $r^{-6}$ scaling (fitted
exponent $-5.87$), tracks the Lorentzian $J(\omega_0)$ as the transition
frequency is tuned (simulated ratios $1.00/0.90/0.68$ against predicted
$1.00/0.90/0.67$), and agrees with the angle-averaged analytic rate to
within the expected $\mathcal{O}(1)$ geometric factor. Second, a direct
three-dimensional evaluation of the bath integral reproduces the closed-form
relaxivity of Eq.~\eqref{eq:kappa} to four significant figures, and
promoting the constant-$\tau_c$ integral to the distance-dependent form of
Eq.~\eqref{eq:tauc}, a translational-diffusion
(Hwang--Freed-type~\cite{hwangfreed1975}) refinement, changes the result by
less than two percent. Third, applying the same relaxivity to NV parameters
($\Tone^{(0)}\sim\SI{3}{ms}$, closest approach $\sim\SI{3}{nm}$,
$\tau_c\sim\SI{0.5}{ns}$) predicts that freely diffusing spin-$\tfrac12$
radicals shorten the NV $\Tone$ measurably at high-micromolar to millimolar
concentrations, consistent with NV free-radical relaxometry reaching
micromolar sensitivity only with large-moment probes such as
$\mathrm{Gd}^{3+}$~\cite{steinert2013,peronamartinez2020,sigaeva2022}. These
checks establish that the relaxation model is quantitatively correct before
it is applied to the FPSQ.

\section{Intrinsic $\Tone$: phonon-limited relaxation and its
engineering}\label{sec:phonon}

The detection limit derived in Sec.~\ref{sec:framework} is directly
proportional to the intrinsic, radical-free relaxation rate
$\Gamma_0 = 1/\Tone$ of the sensor itself. This section establishes what
sets that rate, why it is so large at physiological temperature, and the
central quantitative question for the platform, what would have to change
for it to be small enough.

\subsection{Direct and Raman processes}\label{sec:phononproc}

Reference~\cite{feder2025fluorescent} reports $\Tone = \SI{141(5)}{\micro\second}$ at
\SI{80}{K} together with the temperature dependence
\begin{equation}
\frac{1}{\Tone(T)} = A\,T + B\,T^{7},
\label{eq:AT7}
\end{equation}
with $A = \SI{43(8)}{K^{-1}s^{-1}}$ and $B = \SI{47(7)e-12}{K^{-7}s^{-1}}$.
This is the canonical signature of spin-lattice relaxation in a
non-Kramers spin system, with a one-phonon direct process linear in
$T$ in the high-temperature limit $\hbar\omega_0\ll\kB T$, together with a two-phonon
Raman process whose $T^{7}$ law is the standard non-Kramers
Debye-model result from the $\int x^6 e^x/(e^x-1)^2\,dx$ phonon
integral~\cite{abragam1961,stoneham2001,shrivastava1983}. That a $T^7$ law
fits up to room temperature is itself informative,as it requires
$T\ll\theta_D$ over the fitted range, implying that the responsible modes
have an effective Debye temperature well above \SI{300}{K},stiff
intramolecular vibrations of the chromophore and its hydrogen-bonded pocket
rather than soft global modes of the protein.

\begin{figure*}[t]
\centering
\includegraphics[width=\textwidth]{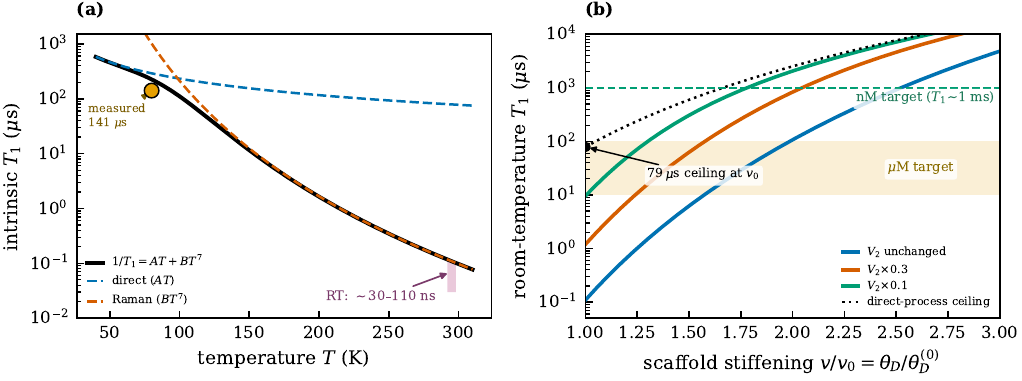}
\caption{\textbf{Phonon-limited intrinsic $\Tone$ and the engineering
target.} (a)
The temperature dependence of the intrinsic $T_1$. The total relaxation rate has contributions from two processes: the direct process (blue line) and the Raman process (orange line), while the black line shows the total relaxation rate. The measured $T_1$ at 80 K (141 $\mu$s) differs from the value obtained from the fitted coefficients (226 $\mu$s) by a factor of $\approx$ 1.6. Extrapolating to room temperature gives a $T_1$ range of $\approx$ 30-110 ns (shaded). At 295 K, 
the Raman contribution is about 720 times larger than the direct contribution and thus dominates the relaxation.
 (b) 
Room-temperature $\Tone$ achievable by making the chromophore environment stiffer and/or by reducing the two-phonon coupling $V_2$.Since $B$ decreases rapidly with increasing stiffness as $v^{-10}$, increasing the stiffness by a factor of about 2 can bring $\Tone$ into the microsecond-sensing range (\SIrange{10}{100}{\micro\second}, shaded), while a factor of about 2.5 can reach the nanomolar target (\SI{1}{ms}, dashed). However, reducing $V_2$ alone cannot overcome the limit set by the direct process (dotted), which is \SI{79}{\micro\second} at the native stiffness.}

\label{fig:phonon}
\end{figure*}

Decomposing Eq.~\eqref{eq:AT7} at the two relevant temperatures, 
shows the relative contributions of the two processes [Fig.~\ref{fig:phonon}(a)]. At \SI{80}{K} the two processes are
comparable, the Raman term contributing only $\sim\!30\%$ of the total,
which is why the low-temperature $\Tone$ is long. At \SI{295}{K} the steep
$T^7$ term overwhelms the direct term by
\begin{equation}
\frac{B\,T^7}{A\,T}\bigg|_{T=\SI{295}{K}}
= \frac{\SI{9.14e6}{s^{-1}}}{\SI{1.27e4}{s^{-1}}} \approx 720 ,
\label{eq:ramandom}
\end{equation}
driving $\Gamma_0$ to $\sim\!\SI{9.1e6}{s^{-1}}$ and the intrinsic $\Tone$
to $\sim\!\SI{0.1}{\micro\second}$. Room-temperature relaxation of
this qubit is thus almost entirely a two-phonon Raman process; any
engineering effort that does not target the Raman channel addresses
$0.14\%$ of the problem.

We note explicitly, as Ref.~\cite{feder2025fluorescent} does, that the fitted
coefficients and the directly measured \SI{80}{K} value are mutually
consistent only to within $\sim\!1.6$, with Eq.~\eqref{eq:AT7} extrapolates to
$\Tone(\SI{80}{K})=\SI{226}{\micro\second}$ against the measured
\SI{141}{\micro\second}. This lies within the large stated uncertainties on
$A$ and $B$ but propagates to the room-temperature extrapolation, which
spans $\Tone\approx\SIrange{30}{110}{ns}$ depending on the anchor
[Fig.~\ref{fig:phonon}(a)]. Because this value is decisive and not directly
measured, we treat the intrinsic $\Tone$ as the primary swept parameter of
the analysis (Sec.~\ref{sec:results}) and hang no conclusion on the precise
extrapolation. The qualitative conclusion that room-temperature relaxation
is Raman-dominated is robust, since the $720\!:\!1$ ratio remains $\gg\!1$
even under a factor-of-two revision of either coefficient.

\subsection{Scaling with stiffness and vibronic coupling}
\label{sec:scaling}

To convert the required $\Tone$ into a materials target we use the standard
Debye-model expressions. Writing $V_1$ and $V_2$ for the one- and two-phonon
spin-phonon coupling matrix elements (linear and quadratic derivatives of
the spin Hamiltonian with respect to the relevant normal coordinates),
$\rho$ for mass density, and $v$ for sound
velocity~\cite{abragam1961,stoneham2001,shrivastava1983},
\begin{equation}
A \;\sim\; \frac{V_1^2\,\omega_0^2}{\rho\,v^5},
\qquad
B \;\sim\; \frac{V_2^2}{\rho^2\,v^{10}} .
\label{eq:scaling}
\end{equation}
The $v^{-5}$ and $v^{-10}$ dependences arise from the one- and two-phonon
densities of states, and since $\theta_D \propto v$ at fixed number density,
they are equivalently $A\propto\theta_D^{-5}$ and $B\propto\theta_D^{-10}$.
Stiffening therefore suppresses both channels, the dominant Raman
channel far more steeply.

This steep scaling is the paper's central mechanistic point and it
transforms the outlook. Taken at face value, raising the room-temperature
$\Tone$ from $\sim\!\SI{0.1}{\micro\second}$ to
\SIrange{10}{1000}{\micro\second}, three to four orders of
magnitude sounds prohibitive. But because $B\propto v^{-10}$, a factor of
two in stiffness alone changes $B$ by $2^{10}\approx10^3$. Evaluating
Eqs.~\eqref{eq:AT7}--\eqref{eq:scaling} at \SI{295}{K}
[Fig.~\ref{fig:phonon}(b)]: $v/v_0 = 1.5$ gives
$\Tone\approx\SI{6}{\micro\second}$ ($\times57$); $v/v_0 = 2.0$ gives
$\Tone\approx\SI{107}{\micro\second}$ ($\times980$), crossing the
micromolar-sensing threshold; and $v/v_0 = 2.5$ gives
$\Tone\approx\SI{0.9}{ms}$ ($\times8400$), reaching the nanomolar target.
Reducing the vibronic coupling instead, a factor of ten in $V_2$ buys
$\times88$. The requirement is therefore not a vague plea for ``better
coherence" but a specific, bounded, falsifiable target: a $\sim\!2\times$
increase in effective stiffness, or an order-of-magnitude reduction in
quadratic vibronic coupling, or a combination.

\subsection{The direct-process ceiling}\label{sec:ceiling}

The two routes are not interchangeable, and this asymmetry yields the most
actionable result of this work. Suppose the Raman channel were eliminated
entirely ($V_2\to0$ at fixed stiffness, perfect vibronic decoupling). The
intrinsic rate would not vanish but saturate at the direct-process floor,
\begin{equation}
\Tone^{\rm ceiling}(T) = \frac{1}{A\,T}
\;\xrightarrow{\;T=\SI{295}{K}\;}\; \SI{79}{\micro\second},
\label{eq:ceiling}
\end{equation}
using the measured $A$. This ceiling is marked in Fig.~\ref{fig:phonon}(b),
where the $V_2$-reduction curves bend over and asymptote to it. Its
consequences are sharp: (i) vibronic decoupling alone can reach the
micromolar window, since \SI{79}{\micro\second} lies within the required
\SIrange{10}{100}{\micro\second}, so detecting micromolar $\NO$ bursts near
active nNOS is compatible with a Raman-suppression-only strategy; (ii)
vibronic decoupling alone \emph{cannot} reach the nanomolar resting-level
target ($\Tone\sim\SI{1}{ms}$), blocked by a factor of $13$ against
Eq.~\eqref{eq:ceiling}; (iii) breaking the ceiling requires reducing $A$
itself, i.e.\ stiffening the lattice ($A\propto v^{-5}$) or reducing the
linear coupling $V_1$. Stiffening is doubly effective because it attacks
both channels, and is the only single lever that reaches the nanomolar
target. The design rule is therefore engineer stiffness, not just
decoupling. A directed-evolution campaign should select for a rigidified
chromophore pocket though tighter hydrogen-bonding networks, reduced local free
volume and suppressed low-frequency torsional modes rather than simply
reduced vibronic coupling, because only the former raises the ceiling.

We stress the limits of this analysis. The Debye model is a crude
description of a protein, whose vibrational spectrum is a discrete set of
localized modes rather than an isotropic continuum, and $v/v_0$ should be
read as an effective stiffness of the modes that dominate the spin--phonon
matrix elements rather than a literal bulk sound velocity.
Eq.~\eqref{eq:scaling} yields scaling guidance, not a first-principles
prediction, and the numbers inherit the factor-$\sim\!1.6$ uncertainty in
$A$ and $B$. What survives is the structural conclusion: room-temperature
relaxation is Raman-dominated; the Raman coefficient depends on stiffness
with a very high power; the required $\Tone$ improvement therefore
corresponds to a modest rather than extravagant change in the vibrational
environment; and a direct-process ceiling separates the micromolar target
from the nanomolar one. Whether directed evolution can deliver a
$2\times$ stiffening of a chromophore pocket is unknown but it is now a
well-posed question with a number attached, which is what a design theory
should provide.

\section{Detection-limit framework}\label{sec:framework}

\subsection{Shot-noise-limited rate resolution}\label{sec:shotnoise}

Relaxometry estimates $\Gamma_1$ from the decay of the triplet polarization
$P(\tau)=P_0\,e^{-\Gamma_1\tau}$ measured after a variable dark time $\tau$.
Writing $C$ for the spin readout contrast and $N_{\mathrm{tot}}$ for the
total number of detected signal photons, a Fisher-information
analysis of the single-exponential estimator ({See \textcolor{blue}{\href{run:Supplementary_Material.pdf}{Supplementary Material}}}) gives an optimal dark time $\tau^\star=1/(2\Gamma_1)$
and a minimum detectable rate change
\begin{equation}
\delta\Gamma_{\min} = \alpha\,\frac{\Gamma_0}{C\sqrt{N_{\mathrm{tot}}}},
\qquad \alpha = 2\sqrt{e}\approx 3.30,
\label{eq:dgamma}
\end{equation}
where $\Gamma_0$ dominates near threshold. The readout model assumes
photon-shot-noise-limited detection with per-shot polarization variance
$\sigma_P^2 = 1/(C^2 n_{\rm ph})$, appropriate here because the OADF scheme
yields $\ll\!1$ delayed photon per molecule per cycle~\cite{feder2025fluorescent},
deep in the single-shot inefficient limit where photon statistics dominate.
Eq.~\eqref{eq:dgamma} carries the two features that drive everything
below: the fractional resolution improves only as $N_{\rm tot}^{-1/2}$,
while the absolute floor is proportional to $\Gamma_0$,so a sensor
with fast intrinsic relaxation has a proportionally worse floor for
resolving the small extra rate contributed by radicals.

\subsection{Photon budget grounded in photobleaching}\label{sec:budget}

$N_{\mathrm{tot}}$ is not a free parameter. We ground it in the demonstrated
performance of Ref.~\cite{feder2025fluorescent}, which collected \num{0.167} delayed
photons per cycle from $7.7\times10^{8}$ molecules, a detected yield of
\begin{equation}
p_{\rm det}^{(0)} = \frac{0.167}{7.7\times10^{8}}
= \SI{2.2e-10}\ \text{photons molecule}^{-1}\text{cycle}^{-1},
\label{eq:pdet}
\end{equation}
which already contains the triplet yield ($\sim\!0.003$), the OADF
branching, and collection/detection efficiencies. The total budget is
\begin{equation}
N_{\rm tot} = N_{\rm mol}\; p_{\rm det}^{(0)}\; G\; N_{\rm cyc},
\label{eq:Ntot}
\end{equation}
with $N_{\rm mol}$ the number of sensor molecules, $G$ the improvement
factor over the demonstrated yield, and $N_{\rm cyc}$ the number of readout
cycles a molecule survives before bleaching. Reference~\cite{feder2025fluorescent}
identifies two routes to $G$: fluorescence cycling ($\lesssim\!333\times$)
and improved optics ($\lesssim\!810\times$), so $G\lesssim2.7\times10^5$ if
both are realized and $G\sim500$ conservatively. Photobleaching data give
survival from $\sim\!15$ s at high power to $\sim\!2000$ s at low power,
with a cycle time set by $\tau^\star=1/(2\Gamma_0)$, an engineered
$\Tone=\SI{1}{ms}$ sensor executes $N_{\rm cyc}\sim10^4$--$10^6$ cycles in
that window.

Eq.~\eqref{eq:Ntot} converts the abstract $N_{\rm tot}$ axis into
concrete requirements, and the result is sobering. Under the conservative
combination ($G=500$, $N_{\rm cyc}=10^4$) the lifetime yield is only
$Y\equiv p_{\rm det}^{(0)}GN_{\rm cyc}\approx10^{-3}$ detected photons per
molecule, so even $10^7$ sensors,a plausible single-cell complement of a
strongly expressed fluorescent protein give $N_{\rm tot}\approx10^{4}$ and
a detection limit of order $\SI{200}{\micro M}$. Under the optimistic
combination ($G=2.7\times10^{5}$, $N_{\rm cyc}=10^{6}$), $Y\approx59$ and
the same $10^7$ sensors give $N_{\rm tot}\approx6\times10^{8}$ and a limit
$\approx\SI{1}{\micro M}$. We report, alongside every operating point, the
sensor number it implies. This makes the ensemble/resolution tradeoff
explicit (Sec.~\ref{sec:map}) and identifies photon yield as a co-equal
bottleneck to $\Tone$.

\subsection{Minimum detectable concentration}\label{sec:cmin}

Setting the bath-induced rate equal to the noise floor,
$\kappa\,n_{\min} = \delta\Gamma_{\min}$,
\begin{equation}
n_{\min} = \frac{\alpha\,\Gamma_0}
                {C\,\kappa\,\sqrt{N_{\mathrm{tot}}}},
\label{eq:nmin}
\end{equation}
which we convert to a molar concentration. Of the four levers-$\kappa$,
$C$, $N_{\rm tot}$, $\Gamma_0$,the first three are bounded within roughly
an order of magnitude by the platform, while $\Gamma_0$ varies over four
decades across the plausible range of Sec.~\ref{sec:phonon} and enters
linearly. This is why the intrinsic $\Tone$ is the natural principal axis.

\subsection{Temporal resolution}\label{sec:bandwidth}

A concentration limit is incomplete for a dynamic signal. Neural $\NO$ and
ROS transients evolve on tens of milliseconds to
seconds~\cite{garthwaite2008}, and the sensor must accumulate its photons
within that window. Combining Eq.~\eqref{eq:Ntot} with a cycle time
$t_{\rm cyc}=(1+\xi)/(2\Gamma_0)$, where $\xi$ parameterizes
initialization/readout overhead, the detected photon rate is
$\dot{N} = N_{\rm mol}\,p_{\rm det}^{(0)}G/t_{\rm cyc}$ and the budget
available within an integration time $\Delta t$ is
$N_{\rm tot}(\Delta t)=\dot{N}\,\Delta t$. Substituting into
Eq.~\eqref{eq:nmin},
\begin{equation}
c_{\min}(\Delta t) =
\frac{\alpha\,\Gamma_0}{C\,r_1}
\left[\frac{2\,\Gamma_0\,N_{\rm mol}\,p_{\rm det}^{(0)}\,G\,\Delta t}
{1+\xi}\right]^{-1/2}.
\label{eq:cmin_dt}
\end{equation}
This demonstrate the sensitivity bandwidth tradeoff.
Since $c_{\min}\propto\Delta t^{-1/2}$, so ten-fold faster resolution costs
$\sqrt{10}\approx3.2$ in sensitivity. It also carries a non-obvious
dependence on $\Gamma_0$. The explicit factor $\Gamma_0$ is partly offset by
the $\Gamma_0^{1/2}$ in the denominator (a faster-relaxing sensor cycles
faster), leaving $c_{\min}\propto\Gamma_0^{1/2}$ at fixed $\Delta t$ rather
than the $\Gamma_0^{1}$ of the static limit. Lengthening $\Tone$ thus
remains beneficial, but with a square-root rather than linear payoff once
the measurement is bandwidth-limited,a qualitatively different scaling
that, to our knowledge, has not been noted in the relaxometry literature
and which tempers the value of very long $\Tone$ for fast transients.

\section{Results}\label{sec:results}

\subsection{The detection-limit map}\label{sec:map}

\begin{figure*}[t]
\centering
\includegraphics[width=\textwidth]{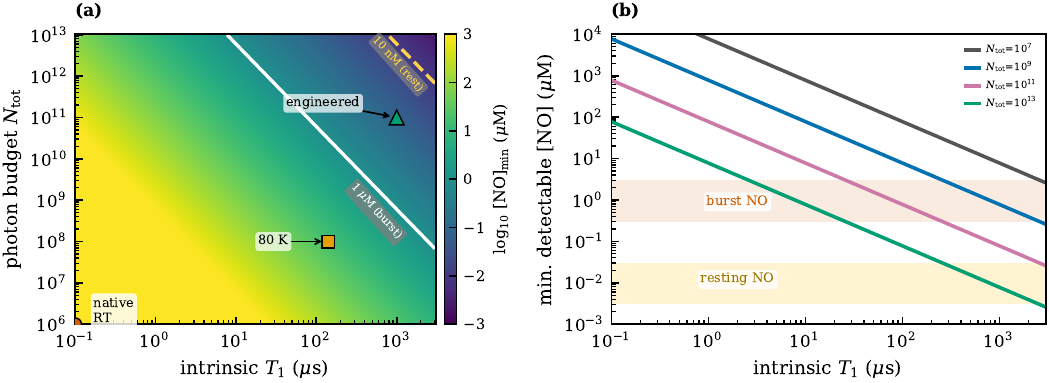}
\caption{\textbf{Detection-limit map.} (a) Minimum detectable $\NO$
concentration (log color scale) versus intrinsic triplet $\Tone$ and
detected-photon budget $N_{\mathrm{tot}}$, for an anchored sensor
($a=\SI{2}{nm}$, $\tau_c=\SI{10}{ps}$, $C=0.20$, corresponding to the lowest ODMR transition).The white solid and yellow dashed lines indicate the micromolar (burst) and nanomolar (resting) sensitivity levels, respectively. The markers show the native room-temperature, 80 K, and engineered operating points.
(b) Minimum detectable NO$^\bullet$ concentration as a function of $T_1$  for fixed photon budgets.The shaded regions indicate  the physiologically relevant NO$^\bullet$ concentration ranges. The limit is
controlled primarily by the intrinsic $\Tone$, where micromolar sensitivity
requires $\Tone\gtrsim\SIrange{10}{100}{\micro\second}$ and nanomolar
sensitivity requires $\Tone\sim\SI{1}{ms}$.}
\label{fig:map}
\end{figure*}

Fig.~\ref{fig:map} is the central result, showing the minimum detectable $\NO$
concentration from Eq.~\eqref{eq:nmin} versus the intrinsic $\Tone$ and the
photon budget $N_{\mathrm{tot}}$, at fixed anchored geometry
($a=\SI{2}{nm}$), nominal $\NO$ correlation time ($\tau_c=\SI{10}{ps}$), the
lowest ODMR transition, and $C=0.20$.The plot also shows the physiological contours for micromolar (burst) and
10~nM (resting) conditions, along with three operating points.
The native room-temperature qubit sits at a detection limit of order
$10^{6}\uM$, roughly six orders of magnitude above even the micromolar burst
level and some eight orders above the resting nanomolar level. Relaxometry
with the qubit as it exists today is hopeless for physiological radicals.
Cryogenic operation with an enhanced photon budget
($\Tone=\SI{141}{\micro\second}$, $N_{\mathrm{tot}}=10^{8}$, $C=0.20$)
reaches the few-tens-of-micromolar range, and a hypothetical engineered
qubit ($\Tone\sim\SI{1}{ms}$, $N_{\mathrm{tot}}=10^{11}$) crosses into the
physiological burst regime.

Because $n_{\min}\propto\Gamma_0/\sqrt{N_{\mathrm{tot}}}$, a decade of
intrinsic $\Tone$ buys a decade of sensitivity whereas a decade of photon
budget buys only half a decade [Fig.~\ref{fig:map}(b)]. The intrinsic
relaxation time is the efficient lever. The qualitative thresholds are
unambiguous micromolar $\NO$ requires $\Tone\sim\SIrange{10}{100}{\micro
\second}$ and resting nanomolar $\NO$ requires $\Tone$ approaching a
millisecond with $N_{\rm tot}\sim10^{12}$. These are the design
specifications the theory yields, and they connect directly to the phonon
targets of Sec.~\ref{sec:phonon}.

\subsection{Temporal resolution and parameter robustness}\label{sec:robust}

\begin{figure*}[t]
\centering
\includegraphics[width=\textwidth]{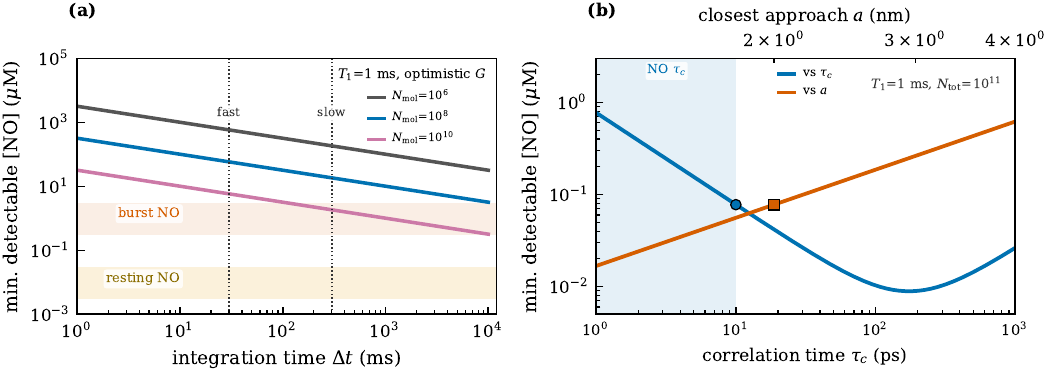}
\caption{\textbf{Temporal resolution and parameter robustness.} (a) Minimum
detectable $\NO$ versus integration time $\Delta t$ ,for $T_1=\SI{1}{ms}$ and an optimistic photon yield, shown for three different numbers of sensor molecules.
The physiological NO concentration ranges are shaded,  the fast and slow transient timescales are marked. The minimum detectable concentration follows $c_{\min}\propto\Delta t^{-1/2}$, showing the tradeoff between sensitivity and temporal resolution.
(b) Static detection limit versus the two most uncertain parameters: the
correlation time $\tau_c$ (bottom axis; $c_{\min}\propto\tau_c^{-1}$, with
the $\NO$ range shaded) and the closest approach $a$ (top axis;
$c_{\min}\propto a^{3}$). Markers show the nominal operating point.}
\label{fig:bandwidth}
\end{figure*}

Fig.~\ref{fig:bandwidth}(a) uses Eq.~\eqref{eq:cmin_dt} to the dynamic problem. Using the demonstrated per-molecule yield with the optimistic
improvement factor, an engineered $\Tone=\SI{1}{ms}$ sensor resolving a
$\Delta t=\SI{100}{ms}$ transient reaches the micromolar burst level only
with $\sim\!10^{8}$ sensors, and reaching resting nanomolar levels at that
bandwidth is out of range for any plausible sensor number. This is the
concrete cost of the sensitivity-bandwidth tradeoff. The static thresholds
of Fig.~\ref{fig:map} are achievable only by integrating over seconds, and
tracking fast physiological transients sacrifices one to two orders of
magnitude in concentration sensitivity. A relaxometric FPSQ is therefore
better suited to reporting sustained or slowly varying radical levels than
to resolving millisecond spikes.

Fig.~\ref{fig:bandwidth}(b) addresses robustness to the two least
certain inputs. The limit scales as $c_{\min}\propto\tau_c^{-1}$ (white-noise
regime) and $c_{\min}\propto a^{3}$, so the order-of-magnitude uncertainty
in the $\NO$ correlation time translates directly into an order of magnitude
in sensitivity, and the choice of closest approach between $1.5$ and
\SI{3}{nm} spans a factor of $\sim\!8$. These dependences are smooth and
monotonic. None introduces a threshold or divergence within the plausible
range, so the qualitative conclusions of Fig.~\ref{fig:map} are robust while
the precise numerical limit inherits these bounded uncertainties. The
$a^{-3}$ growth of the relaxivity as $a\to0$ is cut off physically at
$\sim\!1$ nm by the chromophore spin-density extent and the weak-coupling
condition (Sec.~\ref{sec:bath}), so smaller closest approach cannot be
invoked to rescue sensitivity.

\subsection{The genetic-anchoring advantage and its limit}\label{sec:anchor}

\begin{figure*}[t]
\centering
\includegraphics[width=\textwidth]{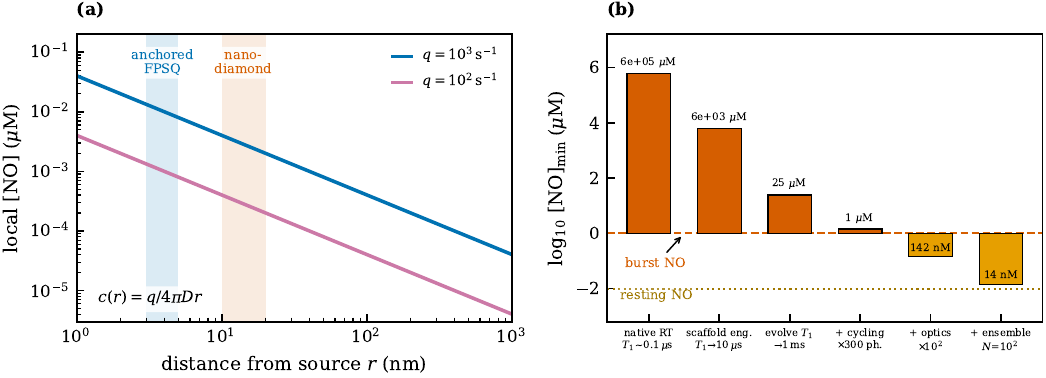}
\caption{\textbf{Anchoring advantage and improvement path.} (a) The steady-state local NO$^\bullet$ concentration is shown as a function of the distance $r$ from a point NO$^\bullet$ source, for two different source strengths. The shaded regions indicate the approximate standoff ranges of the anchored FPSQ and nanodiamond. Even at a distance of 3 nm, a single enzyme producing NO$^\bullet$ at $q\sim10^2-10^3\,\mathrm{s}^{-1}$ sustains only a nanomolar-scale local NO$^\bullet$ concentration.
(b) Improvement waterfall for the minimum detectable NO$^\bullet$ ,
starting from the native room-temperature qubit and successively adding
scaffold engineering of the intrinsic $\Tone$, photon recycling, improved
optics, and ensemble averaging. Reaching the physiological range requires
the full stack, with the intrinsic-$\Tone$ steps carrying most of the
improvement.}
\label{fig:anchor}
\end{figure*}

The distinctive asset of a genetically encoded sensor is placement: fused to
a radical-producing enzyme, the qubit sits within a few nanometers of the
source and samples a local concentration far above the bulk average.
Modeling the source as a point emitter of $q$ radicals per second into an
infinite diffusive medium gives the steady-state profile $c(r)=q/(4\pi D
r)$ [Fig.~\ref{fig:anchor}(a)]. An anchored sensor at \SIrange{3}{5}{nm}
samples a local $\NO$ concentration several-fold higher than a nanodiamond
at \SIrange{10}{20}{nm}, amplified further by the $r^{-6}$ weighting and the
shorter near-field correlation time. This is the concrete sense in which
genetic targeting earns its keep.

The same profile bounds the ambition honestly. A single nNOS enzyme turning
over at $q\sim\SIrange{1e2}{1e3}{s^{-1}}$ sustains a local $\NO$
concentration of only $\sim\!\SI{1}{nM}$ at \SI{3}{nm}
[Fig.~\ref{fig:anchor}(a)], because rapid diffusion disperses the radical
almost as fast as it is produced. Detecting the output of one enzyme
molecule therefore sits at the edge of what even an ideal long-$\Tone$
sensor could achieve, and realistic single-source detection will rely on
bursts of correlated production, microdomains where radical concentration is
locally elevated, or ensemble readout across many anchored sensors which
trades spatial resolution for sensitivity, as the $N_{\rm mol}$ dependence
of the photon budget makes explicit. Fig.~\ref{fig:anchor}(b) collects the
available improvements into a waterfall, showing that no single
improvement suffices and the intrinsic-$\Tone$ steps are both the largest
and the least established.

\subsection{Frequency-resolved species discrimination}\label{sec:discrim}

\begin{figure*}[t]
\centering
\includegraphics[width=\textwidth]{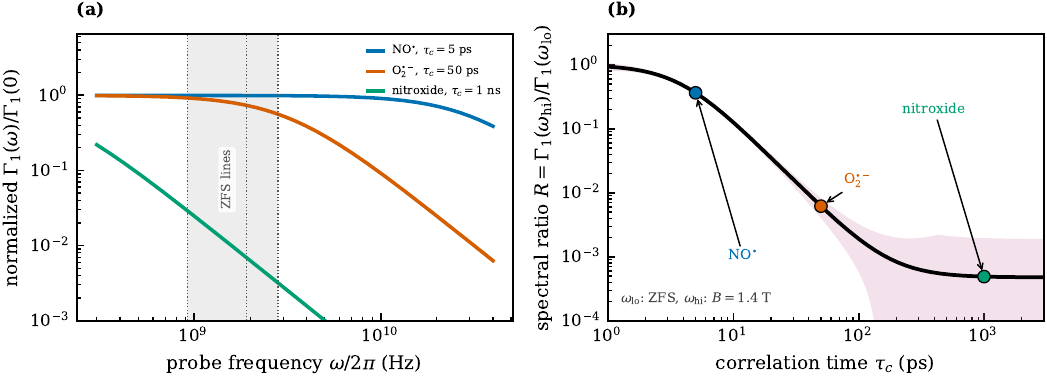}
\caption{\textbf{Frequency-resolved species discrimination.} (a) Normalized
noise spectral density for correlation times representative of $\NO$
(\SI{5}{ps}), superoxide (\SI{50}{ps}), and a nitroxide (\SI{1}{ns}). The
zero-field ODMR window and the field-tuned probe at $B=\SI{1.4}{T}$ are
indicated. (b) Amplitude-independent spectral ratio
$R=\Gamma_1(\omega_{\mathrm{hi}})/\Gamma_1(\omega_{\mathrm{lo}})$ versus
correlation time, using $\omega_{\rm hi}$ from full diagonalization of
$H_0$+Zeeman at \SI{1.4}{T} (the $|0\rangle\!\to\!|{+}1\rangle$ transition,
$\sim\!\SI{42}{GHz}$). The unknown relaxivity concentration prefactor
cancels, so $R$ depends only on $\tau_c$, allowing the three species occupy
well-separated values. The shaded band is the shot-noise uncertainty, which
widens where the high-frequency signal becomes small.}
\label{fig:discrim}
\end{figure*}

Single-frequency relaxometry reports only a rate and cannot say which
radical produced it. The FPSQ offers a spectroscopic handle that
fixed-frequency nanodiamond relaxometry lacks. The rate is proportional to
$J(\omega_0)$, which carries the radical's correlation time through its
corner at $\omega\!\sim\!1/\tau_c$ [Fig.~\ref{fig:discrim}(a)]. By probing
at more than one frequency using the three zero-field ODMR transitions
and, more powerfully, tuning $\omega_0$ well above the ZFS with an applied
field,the correlation time can be read out and the species identified. We
emphasize at the outset that this capability is real but budget-limited: as
Fig.~\ref{fig:discrim}(b) shows, cleanly separating $\NO$ from superoxide
requires both a high applied field and a generous photon budget, whereas
separating fast (\si{ps}) from slow (\si{ns}) radicals is comparatively
easy.

To make the discrimination robust against the unknown
relaxivity-concentration prefactor, we use the ratio of rates at two
frequencies, $R = \Gamma_1(\omega_{\mathrm{hi}})/\Gamma_1(\omega_{\mathrm{lo}})
= J(\omega_{\mathrm{hi}})/J(\omega_{\mathrm{lo}})$, in which the amplitude
$\kappa n$ cancels exactly [Fig.~\ref{fig:discrim}(b)]. Obtaining $\omega_{hi}$ correctly requires diagonalizing the full spin-1 Hamiltonian $H_0$, which includes both the zero-field splitting and Zeeman interaction, rather than simply assuming $|D+E+\gamma_e B|$. At
$B=\SI{1.4}{T}$ the Zeeman energy ($\gamma_e B/2\pi\approx\SI{39}{GHz}$)
exceeds the ZFS, strongly mixing the eigenstates, and the drivable
$|0\rangle\!\to\!|{+}1\rangle$ transition lies at $\approx\SI{42}{GHz}$.
With $\omega_{\rm lo}$ a zero-field ODMR line and this field-tuned
$\omega_{\rm hi}$, the three species map to spectral ratios spanning nearly
two decades, so a measurement of $R$ identifies the correlation-time class
directly. The honest limitation, shown as the shaded band, is photonic. For
very short $\tau_c$ the high-frequency rate is itself small because $J$ has
begun to roll off, so resolving it against the shot-noise floor costs photon
budget, and the ratio degrades where that signal falls below the noise.
Frequency-resolved relaxometry thus delivers a genuine, if budget-limited,
chemical specificity that single-frequency schemes cannot.

\section{Discussion}\label{sec:discussion}

\subsection{Feasibility verdict}\label{sec:verdict}

The analysis yields a clear, quantitative verdict. As it exists today the
room-temperature FPSQ cannot detect physiological neural radicals by
relaxometry. Its detection limit is roughly six orders of magnitude above
the micromolar burst level near active nNOS (and some eight orders above
resting nanomolar levels), and the gap is dominated not by geometry, photon
statistics, or contrast but by the qubit's own phonon-limited intrinsic
relaxation. The steep $T^7$ Raman term drives the room-temperature rate to
$\sim\!10^{7}\,\si{s^{-1}}$, and Eq.~\eqref{eq:nmin} shows the detection
limit is directly proportional to it. Micromolar sensitivity requires an
intrinsic $\Tone\sim\SIrange{10}{100}{\micro\second}$ and nanomolar
sensitivity $\Tone\sim\SI{1}{ms}$, in each case together with a photon
budget that is demanding but given the fluorescence-cycling and optical
gains already identified, not obviously out of reach.

\subsection{Comparison with other modalities}\label{sec:compare}

Table~\ref{tab:compare} places the FPSQ against the established
NV-nanodiamond platform and against the two dominant non-magnetic $\NO$
sensing modalities. The comparison clarifies where the approach is most advantageous. The NV center
enjoys a room-temperature $\Tone$ of milliseconds,three to four orders of
magnitude longer than the FPSQ and a correspondingly lower relaxometry
noise floor, but it cannot place itself: it arrives as an exogenous crystal,
cannot be fused to a chosen protein, and sits \SIrange{10}{20}{nm} from its
target. Electrochemical microelectrodes reach nanomolar $\NO$ with
millisecond response but are micron-scale probes that cannot access
intracellular molecular sites. Genetically encoded fluorescent indicators
are genetically targetable but report indirectly through a
conformational/binding change rather than the radical's paramagnetism, and
are not quantitative for the reactive, short-lived species. The FPSQ
occupies the otherwise-empty corner of the design space. It offers genetic
encodability and nanometer placement at the source, an intracellular vantage
with tissue-penetrating \SI{912}{nm} readout, direct transduction of the
radical's spin, and evolvability at the cost, quantified here, of a far
shorter intrinsic $\Tone$. It is suited to primary-signal transduction at
molecular distance, where placement is decisive and the signal is locally
concentrated, rather than to competition with NV magnetometers in the
far-field high-sensitivity regime. The frequency-resolved discrimination of
Sec.~\ref{sec:discrim} is, moreover, a capability the single-transition NV
does not naturally possess.

\begin{table*}[t]
\caption{Comparison of the anchored FPSQ relaxometry concept with
established modalities for local $\NO$/ROS sensing. FPSQ entries in the
``engineered'' rows assume the intrinsic-$\Tone$ target of
Sec.~\ref{sec:phonon} is achieved,whereas the native-qubit values are substantially worse.}
\label{tab:compare}
\begin{ruledtabular}
\begin{tabular}{lcccc}
 & FPSQ relaxometry & NV nanodiamond & Electrochemical & Fluorescent \\
 & (this work) & relaxometry & microelectrode & indicator (GECI-type) \\
\colrule
Sensing mechanism & radical spin ($T_1$) & radical spin ($T_1$)
   & redox current & binding/conformation \\
Genetic targeting & yes & no & no & yes \\
Source standoff & \SIrange{2}{5}{nm} & \SIrange{10}{20}{nm}
   & \si{\micro\meter} & \SIrange{2}{5}{nm} \\
Intrinsic $\Tone$ (RT) & \SIrange{0.03}{0.1}{\micro\second} (native)
   & \SIrange{1}{3}{ms} & --- & --- \\
Concentration limit & $\sim\!\si{\micro M}$ (engineered)
   & high \si{\micro M}--\si{mM} & \si{nM} & \si{nM}--\si{\micro M} \\
Temporal resolution & \si{ms}--\si{s} (budget-limited) & \si{ms}--\si{s}
   & \si{ms} & \si{ms}--\si{s} \\
Species specificity & yes (spectral ratio) & limited & partial & yes (by design) \\
Intracellular & yes & partial & no & yes \\
\end{tabular}
\end{ruledtabular}
\end{table*}

\subsection{Experimentally testable predictions}\label{sec:predictions}

The theory makes several predictions testable without any speculative
engineering, using purified protein and standard relaxometry. First, a
specific $1/\Tone$-versus-concentration calibration with slope $\kappa$
[Eq.~\eqref{eq:kappa}], measurable by incubating purified FPSQ with known
$\NO$-donor concentrations. Second, a definite ratio between the relaxation
enhancement seen by an FPSQ fused to a radical source and by a free
cytosolic FPSQ, testing the anchoring advantage directly. Third, and most
consequentially, the intrinsic-$\Tone$ threshold, which the theory defines exactly. It states,
how long the room-temperature $\Tone$ must be made before cellular detection
of a given radical concentration becomes feasible, and therefore how much
$\Tone$-directed evolution is warranted before cell experiments are
attempted. Fourth, the phonon prediction is directly falsifiable: if the
$T^7$ law and the Debye scaling hold, then a scaffold mutation that raises
the effective stiffness should lengthen the room-temperature $\Tone$ with
the steep $v^{-10}$ dependence, and the direct-process ceiling of
Eq.~\eqref{eq:ceiling} should manifest as a saturation of $\Tone$ at
$\sim\!\SI{79}{\micro\second}$ once the Raman channel is suppressed. Fifth,
the frequency-resolved spectral ratios $R(\tau_c)$ of
Fig.~\ref{fig:discrim}(b) for $\NO$ versus superoxide are testable by
field-dependent relaxometry on radical-containing solutions.

\subsection{Approximations and limitations}\label{sec:limits}

The analysis rests on controlled approximations, stated plainly.
(i)~\emph{Weak coupling}: the Redfield form assumes $b\tau_c\ll1$, very well
satisfied for diffusing radicals but not for a radical bound in contact with
the chromophore, where a non-perturbative treatment would be needed.
(ii)~\emph{Spin-1 reduction}: we showed the single-exponential description
is exact in the white-noise limit relevant to the target analytes
(Sec.~\ref{sec:singleexp}), for slow radicals the full rate matrix is
required and we use it. (iii)~\emph{Constant correlation time}: the
closed-form relaxivity uses a distance-independent $\tau_c$. The
translational-diffusion refinement changes the result by under two percent
here. (iv)~\emph{Correlation-time uncertainty}: the effective $\tau_c$ of
$\NO$ spans \SIrange{1}{10}{ps}, we sweep it and report its linear effect
(Sec.~\ref{sec:robust}). (v)~\emph{Debye model for a protein}: the phonon
scaling of Sec.~\ref{sec:phonon} is order-of-magnitude guidance, not a
first-principles prediction, and $v/v_0$ is an effective stiffness of the
dominant modes. (vi)~\emph{Room-temperature $\Tone$}: the single most
important number is not directly measured and carries a factor-$\sim\!1.6$
uncertainty, we make it the swept parameter and draw no conclusion depending
on its precise value. None of these approximations flatters the sensor. The
framework is conservative by construction, and its verdict is a lower bound
on the required engineering rather than an optimistic projection.

\section{Conclusion}\label{sec:conclusion}

We have developed and numerically validated a detection-limit theory for
spin relaxometry of neural radicals with the genetically encoded
fluorescent-protein spin qubit, and used it to translate a biosensing challenge into a spin–phonon materials problem with a quantitative answer. The
native room-temperature qubit is far by six to eight orders of
magnitude from physiological sensitivity, and the gap is dominated by the
qubit's own phonon-limited intrinsic relaxation. Decomposing that relaxation
shows it to be overwhelmingly a two-phonon Raman process at room
temperature, so that the required $10^3$--$10^4$ improvement in $\Tone$
corresponds, through the $v^{-10}$ Debye scaling, to a modest
$\sim\!2\times$ stiffening of the chromophore environment; a direct-process
ceiling of $\sim\!\SI{79}{\micro\second}$ then separates the micromolar
target, reachable by vibronic decoupling, from the nanomolar target, which
requires genuine stiffening. Photon yield per molecule emerges as a co-equal
bottleneck, and finite measurement bandwidth imposes a
sensitivity-time-resolution tradeoff with an unusual $\Gamma_0^{1/2}$
scaling. The main message is that FPSQ relaxometry of neural radicals is a well-defined spin-engineering problem with a concrete and experimentally testable target. The platform’s unique ability for directed evolution is well suited to achieving this target. We hope the detection-limit map,
the phonon design rules, and the intrinsic-$\Tone$ threshold serve as a
design specification for that effort.

\begin{acknowledgments}
We gratefully acknowledge financial support from the National Quantum Mission (NQM) of the Department of Science and Technology (DST), Government of India, through Grant No. DST/QTC/NQM/QC/2024/1. We acknowledge the use of Claude (Anthropic) for assistance with literature review and manuscript preparation.
\end{acknowledgments}

\bibliography{refs}

\end{document}